\documentclass[10pt,journal]{IEEEtran}
\IEEEoverridecommandlockouts
\usepackage{cite}
\usepackage{amsmath,amssymb,amsfonts}
\usepackage{algorithmic}
\usepackage{algorithm}
\usepackage{graphicx}
\usepackage{textcomp}
\usepackage{xcolor}
\def\BibTeX{{\rm B\kern-.05em{\sc i\kern-.025em b}\kern-.08em
    T\kern-.1667em\lower.7ex\hbox{E}\kern-.125emX}}
\begin{document}

\title{Intelligent Random Access Framework for Massive and Ultra-Reliable Low Latency IoT Communications}
\author{Muhammad~Waleed~Aftab and~Ishtiaq~Ahmad
\thanks{M.~W.~Aftab is with the Electrical Engineering Department, Gomal University, Pakistan, (e-mail: m.waleed.aftab@gmail.com)}
\thanks{I. Ahmad is with the Electrical Engineering Department, Gomal University, Pakistan, (e-mail: ishtiaqahmad@gu.edu.pk)}}

\maketitle

\begin{abstract}
The Internet of Things (IoT) enables smart cities to achieve the vision of connecting everything by smartly linking gadgets without the need for human interaction. However, due to the rapid proliferation of IoT devices, the amount of data produced accounts for a significant share of all communication services. Hence, 6G-enabled specifications for wireless networks are required to enable both massive and ultra-reliable low-latency access to such a hybrid IoT network. In this article, we propose a smart hybrid random access (SH-RA) scheme for massive connections and ultra-reliable low-latency access in the network architecture of IoT communications. According to numerical results, compared to the other baseline schemes, the SH-RA framework enormously enhances the total access probability and fulfills the quality-of-service (QoS) requirements.
\end{abstract}


\section{Introduction}
\hspace*{3.5mm} \IEEEPARstart{W}{ith} the evolution of inventive technologies and urbanization, intelligent cities have emerged as a popular urban development trend that gained a lot of consideration over the last few years. Techniques like cloud computing and the Internet of Things (IoT) are required to actualize the smart city. The IoT devices act as the cornerstone of the intelligent city that is catching growing interest from industries and academics because of its promise applications of intelligently connecting things without the need for human interaction \cite{b1}.
\\The IoT design has stressed the significance of delivering reliable connectivity to deal with the massive access demand, thus proposing a customized random access (RA) technique. Despite the fact that IoT RA is derived from LTE RA, there still are a few anomalies and challenges, not just in the form of greater access requests but also in terms of the environmental conditions where massive IoT devices are expected to function. As a result, the research community is actively formalizing and studying numerous elements of IoT RA. Initial research has led to the development of theoretical performance models that are frequently evaluated by simulations and emphasize the enormous \cite{b2}. 
\\The first and most important phase in the connection between both the BS and the devices is RA. In a typical cellular communication system, the RA mechanism is intended primarily for human-to-human communication with big data transfer, fewer connections, and minimum energy consumption demands. It is required to create an effective RA technique to enable 5G massive communications having smaller data packet transfer, irregular transmission, and diverse Quality of Service (QoS) needs \cite{b3,b4}. IoT communications have recently been supported by artificial intelligence (AI)-based techniques \cite{b5}, \cite{b6}. Under varied numbers of active devices, a machine learning-based RA technique was presented to discover the optimum access barring factors \cite{b5}. 
\\The authors of \cite{b7} suggested a RA system based on long short-term memory (LSTM), in which a deep learning mechanism based on LSTM is used to assign power to devices. Despite this, the aforementioned studies simply looked at the condition where all users have similar priorities, which is incompatible with a heterogeneous IoT network where ultra-reliable low-latency-sensitive IoT devices exist \cite{b8}. 
\\Gartner predicts that more than 30 billion IoT gadgets will be attached to the network \cite{b9} in the coming future. As a result, the volume of data produced by IoT devices makes up a significant component of all link services \cite{b10}. In the 6G-enabled IoT network for smart cities, massive and ultra-reliable low-latency connections will coexist in a system and require a hybrid RA.
\\In this paper, we utilize massive IoT network architecture and proposed a smart hybrid random access (SH-RA) strategy to provide massive access and ultra-reliable low-latency connectivity to such IoT networks. 
\\ \textbf{Contributions:} We propose an SH-RA system for IoT networks to fulfill the latency and reliability access requirements. IoT devices connect to the network using a contention-free procedure (usually two-step), and massive IoT achieve network access via a timing advance (TA)-aided contention-based connectivity method to fulfill the massive network access demand. For estimating the number of active IoT devices, we present the LSTM prediction system. As a result, the BS can dynamically calculate the settings of multi-user detection based on the required dependability, which assures that all operational IoT nodes can connect to the system in only one attempt, meeting the latency requirement.
\\The remainder of this paper is arranged in the following manner. The conventional contention-based random access scheme is given in Section II. The proposed smart hybrid random access for IoT networks is presented in Section III. Section IV contains the numerical results. Finally, we conclude this paper in Section V.
\begin{figure*}
\centerline{\includegraphics[width=0.8\textwidth]{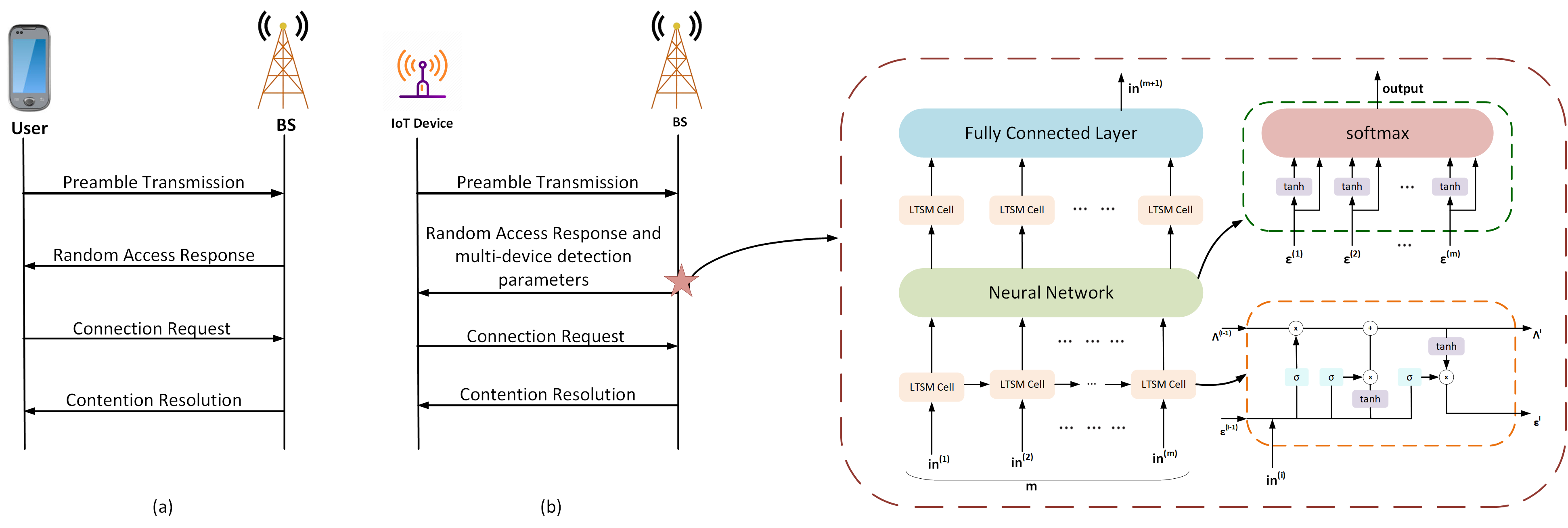}}
\caption{Random Access Framework: (a) Conventional RA Approach (b) Proposed RA Approach.}
\label{fig1}
\end{figure*}
\section{Conventional Contention Based RA Scheme}
RA techniques in the LTE-A cellular connectivity users choose their preambles in a contention-based RA procedure, but non-contention-based RA procedures pre-allocate preambles to some users before the RA operation. The non-contention-driven RA approach is infeasible due to the intermittent transfer of IoT devices. The typical contention-based RA procedure is briefly described in Figure \ref{fig1}(a). The RA procedure is completed successfully if all messages are correctly exchanged. The following are the steps of the conventional RA procedure:
\\ \textbf{\textit{Step 1:}} Each user chooses a preamble using the existing preamble sequences at random and transmits it to the BS through a physical random access channel (PRACH), referred to as the preamble transmission step. Since the preambles sequences are orthogonal, the BS would not be able to distinguish between them if multiple users choose the particular set of preamble in the particular RA slot, resulting in a preamble collision.
\\ \textbf{\textit{Step 2:}} The preamble already received is detected by the BS. The BS might not detect the received preamble if there is preamble conflict. After successful identification of the preamble, the BS sends a Random Access Response (RAR) message along with the detected preamble confirming details and allocates the uplink resource transmits message having connection request.
\\ \textbf{\textit{Step 3:}} The connection request is broadcast on the assigned uplink resource blocks when the IoT device accepts the RAR corresponding to its specified preamble in the particular waiting period. Collisions will occur if numerous devices utilize the same uplink resource block to send the message having a connection request.
\\ \textbf{\textit{Step 4:}} If the BS accepts the message containing the request for connection successfully, it transmits the user a contention resolving message stating the device is successfully connected to the network. The RA process stops when the user obtains this information. The user can then transfer data information to the BS soon after the device, and BS processes a sequence of higher-level signaling.
\\The RA technique is being proposed for human-to-human communication. When it comes to IoT devices, there are some limitations to take into account. Firstly, in comparison to the short data packets transmitted by IoT devices, the above RA approach will impose signaling overhead, lowering data transmission efficiency. Secondly, preamble collision will occur if large massive IoT devices connect to the network. It increases the number of retransmission devices on the network, resulting in increased delay and power consumption. Lastly, because the aforementioned RA process ignores diverse QoS requirements, it is unsuitable for a network with ultra-reliable, low-latency, and massive IoT devices. 
Hence, in 6G-enabled IoT devices, the union of machine learning and broadband connectivity is the leading technique for achieving a smart connection. To address the aforementioned issues, we present an SH-RA scheme based on LSTM models.
\section{Proposed Smart Hybrid Random Access for IoT network}
The proposed SH-RA system is depicted in Figure \ref{fig1}(b), which intends to meet a variety of QoS needs by applying machine learning technology. Furthermore, the cell is partitioned into many annuli with quantized distances from the center to the cell's edge having a radius. Thus, we suppose that every IoT gadget uses length measurement technology to determine its distance from the BS and consequently its TA index. As most devices have fixed locations, so this assumption is plausible. The following are the specifics of the proposed strategy.
\\ The characteristics of the SH-RA approach are given below.
\\ \textbf{\textit{Step 1:}} Every active IoT sensor picks a preamble sequence at random from the set of sequences $\{ \zeta_1, \zeta_2,....\zeta_i...,\zeta_{G} \}$, where $\zeta_i$ represents the Zadoff-Chu sequence with length $D_p$, and $G$ shows the total preambles. Please note that the cyclic prefix is typically included as the head of the preamble sequence to avoid propagation delay. Let $\beta^s_f$ represents the transmitting power of a preamble $s$, $u(f)$ illustrates the index of preamble utilized by IoT node $f$, and $\psi_f*h_f$ shows the channel state information from $f$ IoT node to BS, where $h_f$ is the small scale fading that is distributed between [0,1], and $\psi_f$ is the large scale fading. The IoT devices in the $j^{th}$ annulus use the $[262+j]^{th}$ subcarrier as a preliminary step for their chosen preambles. $\zeta_{s,j}$, $j$ represents the received set of sequences representing a preamble from the $j^th$ annulus, and $k(s,j)$ represents the IoT device in the $j^th$ annulus that selects the preamble.  
\\The preamble signal that is successfully received can thus be expressed as 
\begin{equation}
     r_p = \sum_{j=1}^{\eta} \sum_{f\in F_j} \sqrt{\beta^s_f\psi_f}*h_f* (\zeta_{u(f),j})^T+n_o
\end{equation}
where $n_o$ denotes additive Gaussian noise having a mean of variance $(0,\sigma^2)$. We say $\sqrt{\beta^s_f \psi_f }=1$ which shows that all IoT nodes have the same received power.
\\ \textbf{\textit{Step 2:}} The value of $(s, j), (s = 1,..., \Phi, j = 1,..., \eta)$, can be calculated by the BS using the cross-correlation of $r_p$ and the preamble $\zeta$ with $i^th$ element as shown below: 
\begin{equation}
     y_s^j = [r_p \bigoplus \frac{\zeta_{i}}{||\zeta_{i}||}] = \sqrt{D_p}\sum_{f \in k_{s,j}} h_f + n_o
\end{equation}
The BS then creates a RAR message containing the preamble identity, TA index, and resource block for preamble $s$ with $k_{s,j}=1$. It's worth noting that $k_{s,j}=1$ denotes that in the $j^{th}$ annulus, only one device selects preamble $s$.  As a result, preamble $s$ is associated with numerous RARs, resulting in preamble $s$ being associated with multiple resource blocks. In addition, the BS forecasts total active IoT devices using our suggested LSTM prediction model. As a result, the BS decides and transmits the parameters of multi-devices detection (along with the resource block (particular), modulation, and coding techniques) to active IoT gadgets depending on the reliable transmission requirement.
\\ \textbf{\textit{Step 3:}} Each IoT device searches for RARs that contain its preamble identifier and compares the TA indices in RARs to the one stored by that particular device. If a RAR's TA index matches with an IoT device's TA, it sends a message using the resource block suggested by this RAR with the maximum transmit power level. Conversely, this IoT device selects a particular resource at random from all the resources allotted to the chosen preamble and broadcasts the message at a random transmission power level. Activated IoT devices modulation and coding schemes are utilized on their data to retrieve the messages based on broadcasted parameters, then transmit to BS using the same block from resources.
\\ \textbf{\textit{Step 4:}} The BS uses the successive interference canceller (SIC) technique to decode the received message for every resource carrying a note regarding IoT devices. From the upper to the lower power level, the SIC algorithm analyses the uplink message of devices. The IoT devices with the maximum power, in particular, are being processed immediately. When it detects successfully, the conflict of this device's data statistics is removed, as well as the data of IoT gadget with the 2nd greatest power is taken till the device's data statistics won't detect properly. Consider that there are $A$ distinct levels of power donated by $\{ a_1,a_2,...,a_i,...a_A \}$ that satisfies the condition $a_1>....>a_A$ and $a_i$ can be written as
\begin{equation}
    a_i = \gamma(\gamma+1)^{A-i}
\end{equation}
Here, $\gamma$ presents the signal to interference and noise ratio.
\begin{algorithm}
\scriptsize
\caption{Proposed SH-RA for IoT devices}
\begin{algorithmic} 
\REQUIRE Random Access
\ENSURE Avoid collision
\STATE IoT pick preamble from set $\xrightarrow{}\{ \zeta_1, \zeta_2,....\zeta_i...,\zeta_{G} \}$
\STATE BS Calculates $\sqrt{\beta^s_f\psi_f}*h_f* (\zeta_{u(f),j})^T$
\STATE BS estimates correlation between $r_p$ and $\zeta$
\IF{Correlate}
\STATE BS sends RAR
\STATE Calculate total IoT active nodes using the LSTM model
\STATE RAR message (preamble identity, TA index, resource block, and IoT active nodes).
\IF{$k_{s,j}==1$}
\STATE one IoT node selects $s$
\ENDIF
\STATE Each IoT searches for RAR.
\IF{RAR for IoT}
\STATE IoT sends RA using a resource block
\ENDIF
\STATE BS uses SIC to extract Data
\ENDIF
\end{algorithmic}
\end{algorithm}
\subsection{LSTM prediction Design}
The proposed RA technique uses a prediction model based on LSTM in order to predict the active IoT devices and configures the settings of large detection of devices for IoT devices to fulfill the requirements of low-latency and high-reliability of IoT sensors. As illustrated in Figure \ref{fig2}b, the prediction model that is proposed consists of 2 layers of LSTM, one neural network, and one fully-connected layer. To estimate the operational IoT users, the dataset passes through the first layer of LSTM to the neural network layer, then enters into the second LSTM layer, and eventually links to a fully connected layer. There are numerous LSTM cells in each LSTM layer, and each LSTM cell contains input, output, and forget gates. An LSTM cell, $in = [in(1), in(2),, in(m)]$ represents the LSTM's input data, $m$ is the time step length. For the $i$ time step, the process of creating forget gate $ \lambda_{f}^{i}$, input gate $\Lambda_{c}^{i}$, and output gate $\epsilon^{i}$ is
\begin{equation}
     \lambda_{f}^{i} = \alpha*(b^{(f)}[h ^{(i-1)},in^{i}]*w^{(f)}) 
\end{equation}
\begin{equation}
     \Lambda_{c}^{i} = \tanh*(b^{(c)}[h ^{(i-1)},in^{i}]*w^{(c)}) 
\end{equation}
\begin{equation}
     \epsilon^{i} = \lambda_{f}^{i}*  \Lambda_{c}^{i}
\end{equation}
where $w$ represents the weight and $b$ is the bias. $\alpha$ is used for sigmoid and $\tanh$ for hyperbolic tangent activation functions.
\\ A neural network is used as a helpful tool to boost the effectiveness of LSTM networks since it can combine related information and enable the model to offer dynamic focus to some valuable input information. We suppose that the output of $n^th$ layer during $l^th$ time step is $LT_l^n$, $(n=1,2, l=1,2,...m)$
\begin{equation}
    output^j = \sum_{l=1}^m w^n_l*LT_l^{n-1}
\end{equation}
where $w$ is the weight of the neural network that can be given as
\begin{equation}
    w^n_l = \frac{\exp(e^n_l)}{\sum_{l=1}^m \exp(e^n_l)}
\end{equation}
where $e^n_l$ represents the aggregation function and it is obtained as
\begin{equation}
    e^n_l = Q^T \tanh (w_in^1 [LT_1^{n-1},...,LT_m^{n-1}] + w_in^2 LT_{l-1}^n)
\end{equation}
where, $Q, w_in^1, w_in^2$ are the input weights, and the output of LTSM is $[LT_1^{n-1},...,LT_m^{n-1}]$. Finally, we concatenate $output^j$ and $[LT_1^{n-1},...,LT_m^{n-1}]$ to input it to the second layer of LSTM.
\\The output $in(i+1)$ is retrieved after the features extracted during the last later of LSTM connect to a fully-connected layer. We retrieve the projected active IoT users after rounding the output $in(i+1)$.
\section{Simulation Results and Discussions}
\begin{figure}
\centerline{\includegraphics[width=0.35\textwidth]{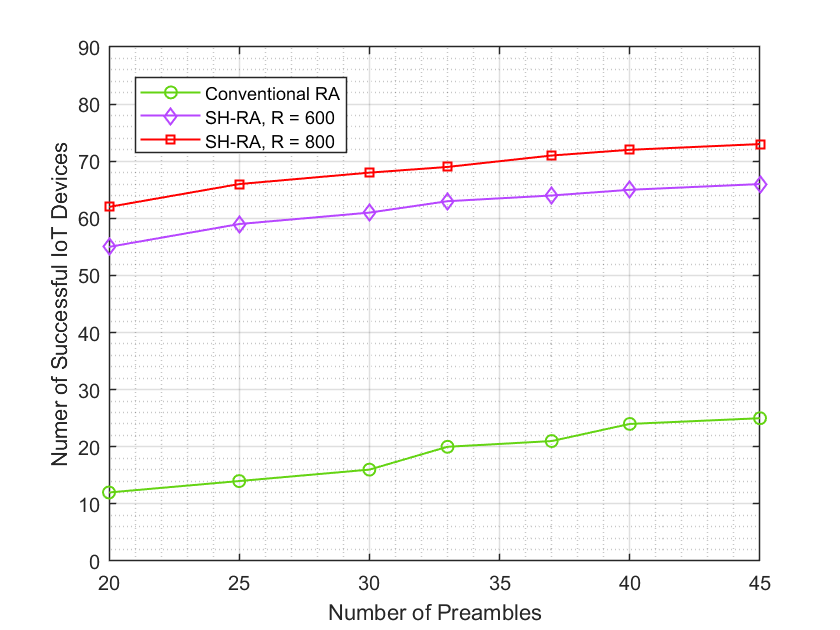}}
\caption{Successful IoT Devices Vs Number of Preambles}
\label{fig2}
\end{figure}

\begin{figure}
\centerline{\includegraphics[width=0.35\textwidth]{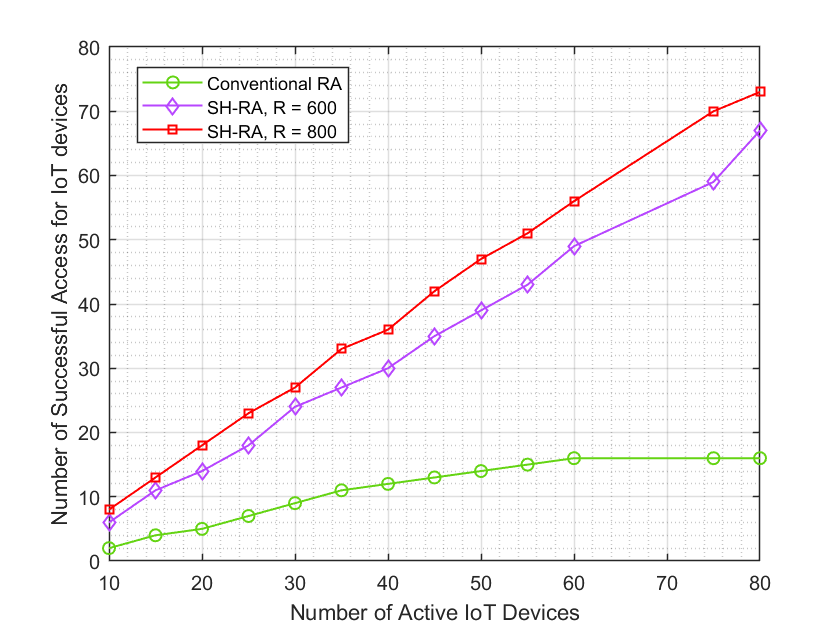}}
\caption{Successfully accessed IoT Devices Vs Total Active IoT devices}
\label{fig3}
\end{figure}

The proposed SH-RA technique aims to satisfy various communication needs of reliability and low latency access of massive IoT nodes. The BS chooses the multi-user detection algorithm's settings based on the predicted total active IoT nodes in order to meet the performance standard. Depending on the volume of active IoT nodes, the modulation and coding techniques can be modified to adjust the requirements of high reliability and low latency. If the predicted IoT nodes are fewer than the actual value, we assume that messages from the IoT nodes cannot be effectively decoded because of inadequate coding and modulation techniques.
\\In proportion to the number of successful access devices, we compare our proposed SH-RA approach to the traditional contention-based random access technique. As a baseline, we utilize a conventional RA framework. In addition, we set the quantization unit to $2*d = 157m$, where $c = 3*10^8m/s$, and $T = 3*10^7s$, respectively. The learning rate is $0.01$ for the LSTM prediction model, and the root-mean-square function is used as a loss function. In each time slot, the total functional IoT devices follow a Poisson distribution with a mean of $3$.
\\The successful IoT users changes with the preambles for radius R = 800, and 600 is shown in Figure \ref{fig2}. The functioning IoT users are set to 80, and the power levels are set to three. As shown in Figure \ref{fig2}, the total successful IoT devices increase as the number of preambles grows and is much greater compared with the baseline technique. It shows that the transmission power allocation technique works well and increases the accessed successful devices.
\\As shown in Figure \ref{fig3}, the total number of successful devices falls slowly for the conventional RA framework and gradually increases for the proposed SH-RA framework as the total of functional IoT devices increases. For 35 active IoT devices, we can also see that number of successful IoT devices connected to the network in the proposed SH-RA scheme is 33 for R=800 and 28 for R=600, which has considerably higher successful devices than the baseline, where only 11 devices are connected.
\section{Conclusions}
We present the hybrid architecture of an IoT network to fulfill the requirements for massive IoT (tremendous connections), and ultra-reliable, low latency access for 6G-enabled IoT devices. The proposed RA framework greatly enhances the successful access probability and satisfies the different QoS demands of IoT devices when compared to the baseline scheme. In the future, we plan to implement and further improve successful massive connection and ultra-reliable low latency access by implementing solutions based on deep learning.

\vspace{12pt}
\color{red}


\begin{thebibliography}{00}
\bibitem{b1} P. Yadav and S. Vishwakarma, “Application of the internet of things and big data towards a smart city,” in 2018 3rd International Conference On Internet of Things: Smart Innovation and Usages (IoT-SIU), 2018, pp. 1–5.
\bibitem{b2} Y. Sun, F. Tong, Z. Zhang, and S. He, “Throughput modeling and analysis of random access in narrowband Internet of Things,” IEEE Internet Things J., vol. 5, no. 3, pp. 1485–1493, Jun. 2018.
\bibitem{b3} M. Agiwal, M. K. Maheshwari, and H. Jin, “Power efficient random access for massive NB-IoT connectivity,” Sensors, vol. 19, no. 22, p. 4944, 2019.
\bibitem{b4} Y. Sun, F. Tong, Z. Zhang, and S. He, “Throughput modeling and analysis of random access in narrowband Internet of Things,” IEEE Internet Things J., vol. 5, no. 3, pp. 1485–1493, Jun. 2018.
\bibitem{b5} T.-O. Luis, P.-P. Diego, P. Vicent, and M.-B. Jorge, “Reinforcement learning-based ACB in LTE-A networks for handling massive M2M and H2H communications,” in Proceeding of 2018 IEEE International Conference on Communications (ICC), 2018, pp. 1–7.
\bibitem{b6} N. Ye, X. Li, H. Yu, A. Wang, W. Liu, and X. Hou, “Deep learning aided grant-free NOMA toward reliable low-latency access in tactile Internet of Things,” IEEE Transactions on Industrial Informatics, vol. 15, no. 5, pp. 2995–3005, 2019.
\bibitem{b7} G. Gui, H. Huang, Y. Song, and H. Sari, “Deep learning for an effective non-orthogonal multiple access scheme,” IEEE Transactions on Vehicular Technology, vol. 67, no. 9, pp. 8440–8450, 2018.
\bibitem{b8} N. H. Mahmood, H. Alves, O. A. L´opez, M. Shehab, D. P. M. Osorio, and M. Latva-Aho, “Six key features of machine type communication in 6G,” in 6G Wireless Summit. IEEE, 2020, pp. 1–5.
\bibitem{b9} Gartner, “Gartner says the Internet of Things will transform the data center,” [Online]. Available: http://www.gartner.com/newsroom/id/2684616, 2014.
\bibitem{b10}H. Aksu, L. Babun, M. Conti, G. Tolomei, and A. S. Uluagac, “Advertising in the IoT era: Vision and challenges,” IEEE Communications Magazine, vol. 56, no. 11, pp. 138–144, 2018.
\end{thebibliography}
\end{document}